\documentclass[11pt]{article}
\usepackage{graphicx}

\newcommand{\BABARPubYear}    {03}

\newcommand{\BABARProcNumber} {067}
\newcommand{\SLACPubNumber} {10127}

\input babarsym

\def\beq{\begin{equation}}
\def\eeq#1{\label{#1}\end{equation}}
\def\eeqn{\end{equation}}

\def\beqa{\begin{eqnarray}}
\def\eeqa#1{\label{#1}\end{eqnarray}}
\def\eeqan{\end{eqnarray}}

\let\bar=\overbar

\def\Dslash{\not{\hbox{\kern-4pt $D$}}}
\def\dslash{\not{\hbox{\kern-2pt $\del$}}}

\def\msb{{\bar{\ssstyle M \kern -1pt S}}}

\def\BB0bar{B^0 {\overline B}^0}
\def\BB0dbar{B_d^0 {\overline B}_d^0}
\def\BB0sbar{B_s^0 {\overline B}_s^0}

\RequirePackage{xspace}

\usepackage{relsize}
\def\babar{\mbox{\slshape B\kern-0.1em{\smaller A}\kern-0.1em
    B\kern-0.1em{\smaller A\kern-0.2em R}}}

\def\en         {\ensuremath{e^-}\xspace}   
\def\ep         {\ensuremath{e^+}\xspace}

\def\mup        {\ensuremath{\mu^+}\xspace}
\def\mun        {\ensuremath{\mu^-}\xspace} 
\def\mumu       {\ensuremath{\mu^+\mu^-}\xspace}

\def\taup       {\ensuremath{\tau^+}\xspace}
\def\taum       {\ensuremath{\tau^-}\xspace}

\def\ellell     {\ensuremath{\ell^+ \ell^-}\xspace}

\def\nub        {\ensuremath{\overline{\nu}}\xspace}
\def\nunub      {\ensuremath{\nu{\overline{\nu}}}\xspace}
\def\nub        {\ensuremath{\overline{\nu}}\xspace}
\def\nunub      {\ensuremath{\nu{\overline{\nu}}}\xspace}
\def\nue        {\ensuremath{\nu_e}\xspace}
\def\nueb       {\ensuremath{\nub_e}\xspace}

\def\num        {\ensuremath{\nu_\mu}\xspace}
\def\numb       {\ensuremath{\nub_\mu}\xspace}

\def\nut        {\ensuremath{\nu_\tau}\xspace}
\def\nutb       {\ensuremath{\nub_\tau}\xspace}

\def\nul        {\ensuremath{\nu_\ell}\xspace}
\def\nulb       {\ensuremath{\nub_\ell}\xspace}

\def\g     {\ensuremath{\gamma}\xspace}
\def\gaga  {\ensuremath{\gamma\gamma}\xspace}  
\def\ggstar{\ensuremath{\gamma\gamma^*}\xspace}

\def\d     {\ensuremath{d}\xspace}

\def\b     {\ensuremath{b}\xspace}

\def\piz   {\ensuremath{\pi^0}\xspace}

\def\ppz   {\ensuremath{\pi^0\pi^0}\xspace}
\def\pip   {\ensuremath{\pi^+}\xspace}
\def\pim   {\ensuremath{\pi^-}\xspace}
\def\pipi  {\ensuremath{\pi^+\pi^-}\xspace}

\def\pimp  {\ensuremath{\pi^\mp}\xspace}

\def\kaon  {\ensuremath{K}\xspace}
\def\Kbar  {\kern 0.2em\overline{\kern -0.2em K}{}\xspace}

\def\Kz    {\ensuremath{K^0}\xspace}
\def\Kzb   {\ensuremath{\Kbar^0}\xspace}
\def\KzKzb {\ensuremath{\Kz \kern -0.16em \Kzb}\xspace}
\def\Kp    {\ensuremath{K^+}\xspace}
\def\Km    {\ensuremath{K^-}\xspace}
\def\Kpm   {\ensuremath{K^\pm}\xspace}

\def\KpKm  {\ensuremath{\Kp \kern -0.16em \Km}\xspace}
\def\KS    {\ensuremath{K^0_{\scriptscriptstyle S}}\xspace} 
\def\KL    {\ensuremath{K^0_{\scriptscriptstyle L}}\xspace} 
\def\Kstarz  {\ensuremath{K^{*0}}\xspace}

\def\Kstar   {\ensuremath{K^*}\xspace}

\def\Kstarp  {\ensuremath{K^{*+}}\xspace}

\def\Dbar    {\kern 0.2em\overline{\kern -0.2em D}{}\xspace}

\def\Dz      {\ensuremath{D^0}\xspace}
\def\Dzb     {\ensuremath{\Dbar^0}\xspace}
\def\DzDzb   {\ensuremath{\Dz {\kern -0.16em \Dzb}}\xspace}
\def\Dp      {\ensuremath{D^+}\xspace}
\def\Dm      {\ensuremath{D^-}\xspace}

\def\DpDm    {\ensuremath{\Dp {\kern -0.16em \Dm}}\xspace}
\def\Dstar   {\ensuremath{D^*}\xspace}

\def\Dstarz  {\ensuremath{D^{*0}}\xspace}

\def\Dstarp  {\ensuremath{D^{*+}}\xspace}
\def\Dstarm  {\ensuremath{D^{*-}}\xspace}

\def\B       {\ensuremath{B}\xspace}
\def\Bbar    {\kern 0.18em\overline{\kern -0.18em B}{}\xspace}

\def\BB      {\ensuremath{B\Bbar}\xspace} 
\def\Bz      {\ensuremath{B^0}\xspace}
\def\Bzb     {\ensuremath{\Bbar^0}\xspace}
\def\BzBzb   {\ensuremath{\Bz {\kern -0.16em \Bzb}}\xspace}
\def\Bu      {\ensuremath{B^+}\xspace}
\def\Bub     {\ensuremath{B^-}\xspace}

\def\Bpm     {\ensuremath{B^\pm}\xspace}

\def\BpBm    {\ensuremath{\Bu {\kern -0.16em \Bub}}\xspace}

\def\jpsi     {\ensuremath{{J\mskip -3mu/\mskip -2mu\psi\mskip 2mu}}\xspace}

\mathchardef\Upsilon="7107
\def\Y#1S{\ensuremath{\Upsilon{(#1S)}}\xspace}

\def\FourS {\Y4S}

\mathchardef\Deltares="7101
\mathchardef\Xi="7104
\mathchardef\Lambda="7103
\mathchardef\Sigma="7106
\mathchardef\Omega="710A

\def\Deltabar{\kern 0.25em\overline{\kern -0.25em \Deltares}{}\xspace}
\def\Lbar{\kern 0.2em\overline{\kern -0.2em\Lambda\kern 0.05em}\kern-0.05em{}\xspace}
\def\Sigbar{\kern 0.2em\overline{\kern -0.2em \Sigma}{}\xspace}
\def\Xibar{\kern 0.2em\overline{\kern -0.2em \Xi}{}\xspace}
\def\Obar{\kern 0.2em\overline{\kern -0.2em \Omega}{}\xspace}
\def\Nbar{\kern 0.2em\overline{\kern -0.2em N}{}\xspace}
\def\Xb{\kern 0.2em\overline{\kern -0.2em X}{}\xspace}

\def\X {\ensuremath{X}\xspace}

\def\BR         {{\ensuremath{\cal B}\xspace}}

\def\bpsikl     {\ensuremath{\Bz \to \jpsi \KL}\xspace}

\def\Bzbtox     {\ensuremath{\Bzb \to \X}\xspace}

\def\de         {\ensuremath {\Delta E^{*}}\xspace}

\def\mes        {\mbox{$m_{\rm ES}$}\xspace}

\def\ebeam      {\ensuremath {E^{*}_\mathrm{beam}}\xspace}

\def\invfb   {\ensuremath{\mbox{\,fb}^{-1}}\xspace}

\def\mus  {\ensuremath{\rm \,\mus}\xspace}

\def\mus        {\ensuremath{\,\mu{\rm s}}\xspace}    

\def\degrees{\ensuremath{^{\circ}}\xspace}

\def\to                 {\ensuremath{\rightarrow}\xspace}

\def\pep2{PEP-II}
\def\BF{$B$ Factory}

\long\def\inst#1{\par\nobreak\kern 4pt\nobreak
    {\it #1}\par\vskip 10pt plus 3pt minus 3pt}

\begin{document}
{\pagestyle{empty}

\begin{flushright}
SLAC-PUB-\SLACPubNumber \\
\babar-PROC-\BABARPubYear/\BABARProcNumber \\
August, 2003 \\
\end{flushright}

\par\vskip 4cm

\begin{center}
\Large \bf Experimental Status of {\boldmath$\b\to s(d) \gamma$} Decays
\end{center}
\bigskip

\begin{center}
\large 
F. Di Lodovico\\
University of Wisconsin \\
1150 University Avenue\\
 Madison, Wisconsin 53706-1390, USA 
\end{center}
\bigskip \bigskip

\begin{center}
\large \bf Abstract
\end{center}
Radiative penguin decays provide an indirect probe 
for physics beyond the Standard Model and contribute to the
determination of the CKM matrix elements. 
Copious quantities of \B\ mesons produced at the \B--Factories permit precision measurements of radiative
penguin decays. We review the experimental status of the
radiative penguin processes $\b\to s(d) \gamma$.

\vfill
\begin{center}
Contributed to the Proceedings of 
Flavor Physics and CP Violation (FPCP 2003)\\
3-6 Jun 2003, Paris, France
\end{center}

\vspace{1.0cm}
\begin{center}
{\em Stanford Linear Accelerator Center, Stanford University, 
Stanford, CA 94309} \\ \vspace{0.1cm}\hrule\vspace{0.1cm}
Work supported in part by Department of Energy contract DE-AC03-76SF00515.
\end{center}

\newpage
\pagestyle{plain}
\section{Introduction}
Radiative penguin decays are flavor--changing neutral current (FCNC)
transitions that are forbidden in the Standard Model
(SM) at the tree level, but occur at the loop level. We will focus 
on the transition with a photon in the final state which happens through penguin loops.
Additional contributions to the electromagnetic 
penguin loops could arise from New Physics effects such as new gauge bosons,
charged Higgs bosons or supersymmetric (SUSY) particles. These interfere with the
SM processes. Depending on the sign of the interference term enhanced or
depleted branching fractions (${\cal B}$'s) result. Moreover, due to the presence of
new weak phases; \CP\ asymmetries that are small in the SM may be enhanced. 
This would make it possible to observe indirectly New Physics from the 
study of the radiative penguin decays.

In addition, these decays are relevant to the determination of the CKM matrix elements.
The measurement of the photon energy spectrum, largely insensitive to New Physics,
can be used to improve the extraction of the CKM elements $V_{cb}$ and $V_{ub}$
from inclusive semileptonic \B\ decays.
Also, a direct estimate of the ratio of the CKM parameters $|V_{td}/V_{ts}|$ can be obtained
from the ratio of the branching fractions for the processes $b\to d\gamma$ and $b\to s\gamma$.

In the following sections, the $b \to s \gamma$ followed by the $b \to d \gamma$ process will be reviewed.
\section{$b \to s \gamma$ final states}
The SM $b\to s\gamma$ branching fraction is 
predicted to be ${\cal B}(b \rightarrow s \gamma) = (3.73 \pm 0.3) \times 10^{-4}$~\cite{misiak}
at the next--to--leading order (NLO). The present theoretical uncertainty of $\sim 10\%$ is dominated by the 
mass ratio of the $c$--quark and $b$--quark and the choice of the renormalization scale. 
New Physics contributions with e.g. charged Higgs exchanges or chargino--squark loops
are at the same level as the SM ones. In the Hamiltonian they could appear as
new operators or new contributions to the coefficients of the SM operators.

\CP asymmetries provide another test of the SM. While small in the SM 
($\leq 1\%$) \cite{soares} they can reach $10$--$50\%$ in models beyond the SM~\cite{kagan}.

For the exclusive decay rate $B \rightarrow K^* \gamma$ 
two recent NLO calculations predict SM branching fractions of 
${\cal B} (B  \rightarrow K^* \gamma) = (7.1^{+2.5}_{-2.3}) \times 10^{-5}$
\cite{bosch} and 
${\cal B} (B  \rightarrow K^* \gamma) = (7.9^{+3.5}_{-3.0}) \times 10^{-5}$
\cite{beneke}. 
The errors are still dominated by the uncertainties in the form factors.

Differently from the branching fraction and \CP\ asymmetry, the photon energy spectrum in
$b\to s\gamma$ is not very sensitive to New Physics processes.
Moments of the photon energy spectrum can be used to measure the Heavy Quark Effective Theory
(HQET) parameters which determine the $b$--quark pole mass ($\overline{\Lambda}$) and the 
kinetic energy ($\lambda_1$)~\cite{ligeti}. 
These parameters are needed to obtain a precision value of
$| V_{cb} |$ from the $b \to c \ell \nu$ inclusive rate.
 
Moreover, the photon spectrum can be used for the determination of $V_{ub}$
from $\B \to X_u \ell \nu$. 
To lowest order in $\Lambda_{QCD}/M_B$, the $B \rightarrow X_s \gamma$ photon energy spectrum,
where $X_s$ refers to inclusive strange hadronic states, 
is given by a convolution of
the parton level $b \to s \gamma$ photon energy spectrum with the
light--cone shape function of the \B\
meson, which describes {\cal all} $b$ to light--quark transitions.
At the same order in $\Lambda_{QCD}/M_B$, 
the $\B \to X_u \ell \nu$ lepton energy spectrum is given by a
convolution of the parton level $b \to u \ell \nu$ lepton energy
spectrum with the same shape function~\cite{vub}.
Corrections enter at next order in $\Lambda_{QCD}/M_B$, and these 
are currently the subject of active investigation~\cite{bauer}.
%
\subsection{The Exclusive Process $\B\to K^*\gamma$}

The exclusive $B \rightarrow K^* \gamma$ modes have been studied by 
\babar~\cite{babar1}, Belle~\cite{belle1} and CLEO~\cite{cleo1}, where
Belle used the highest statistics sample. 
Utilizing kinematic constraints resulting from a full $B$ 
reconstruction in the $B$ rest frame provides
a substantial reduction of the $q \bar q$ (continuum) background.
The Belle \emph{beam--constrained}~\footnote{Results for exclusive \B\ decays 
are typically presented using the following kinematic variables.  If $(E_B^*, \vec{p}_B^*) $
is the four--momentum of a reconstructed \B candidate in the overall
CM (\Y4S) frame, we define

\begin{center}
$ \de \equiv E_B^* - \ebeam \ ,$
 \label{eq:de}
\end{center}
\begin{center}
 $\mes\ (\mathrm{or}\ M_{bc}) \equiv \sqrt{{\ebeam}^2 - p_B^{*2}}\ .$
 \label{eq:mes}
\end{center}
The latter is called the \emph{energy--substituted} (\babar) or
\emph{beam--constrained} (Belle, CLEO) mass.  Signal events peak at \de\ near 0~\gev\
and \mes ($M_{bc}$) near \B\ meson mass; whereas continuum
background lacks peaks.} mass distribution is shown in Figure~\ref{fig:kstar}  for all the 
$K^*$ decay channels.

The measured branching fractions from all the experiments and the corresponding 
average are summarized in Table~\ref{tab:kstar}.
\begin{figure}[hbtp]
\begin{center}
\includegraphics[width=10cm,height=7cm]{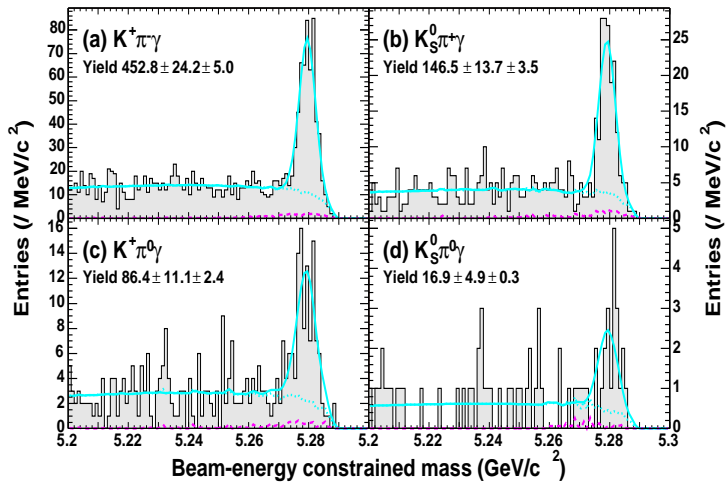}
\caption{Belle~\cite{belle1} \emph{beam--constrained} mass distribution for the 
exclusive $B\to K^*\gamma$ for the four $K^*$ final states.}
\end{center}
\label{fig:kstar}
\end{figure}

Due to the large theoretical errors of $35$--$40\%$, the average branching fraction measurement
is still consistent with the NLO SM predictions.

The direct \CP\ asymmetry is defined, at the quark level, as:
\begin{displaymath}
A_{CP} = {{\Gamma(b\to s \gamma ) - \Gamma(\overline{b}\to \overline{s} \gamma)}\over{\Gamma(b\to s \gamma ) + \Gamma(\overline{b}\to \overline{s} \gamma)}}.
\end{displaymath}
Table~\ref{tab:kstar} summarizes the measurements of the direct \CP\ asymmetry in 
$B\to K^*\gamma$. These are consistent with zero and are statistics limited.

Isospin asymmetry, $\Delta_{0+}$, is calculated by Belle using the world average value 
$\tau_{\B^+}/\tau_{\B^0}$ = 1.083 $\pm$ 0.017~\cite{PDG}. The result is:
$\Delta_{0+} = + 0.003 \pm 0.045 \pm 0.018$, where the first error is statistical and the
second systematic. It is consistent with no asymmetry, having assumed equal production
of charged and neutral \B's at the $\Upsilon(4S)$ mass peak. The isospin asymmetry 
can be used to set limits on Wilson coefficients~\cite{isospin}. 
\hspace{-1cm}
\begin{table}[htbp]
\begin{center}
\begin{tabular}{|l|c|c|c|} \hline
                   & ${\cal B}(B^0\to K^{*0}\gamma )\times10^{-5}$ & ${\cal B}(B^+\to K^{*+}\gamma) \times10^{-5}$ & $A_{CP}$  \\ \hline
\babar~\cite{babar1} (21~\invfb)                 &$4.23\pm 0.40\pm 0.22$&$3.83\pm 0.62\pm 0.22$                    & $-0.044\pm0.076\pm 0.012$\\
Belle~\cite{belle1} (78~\invfb)                  &$4.09\pm 0.21\pm 0.19$&$4.40\pm 0.33\pm 0.24$                    & $-0.001\pm 0.044\pm 0.008$\\
CLEO~\cite{cleo1} (9~\invfb)                     &$4.55{\,}^{+0.72}_{-0.68}\pm 0.34$&$3.76{\,}^{+0.89}_{-0.83}\pm 0.28$  & $+0.08\pm 0.13\pm 0.03$ \\ \hline
Average                                          &$4.18\pm 0.23$ & $4.14\pm 0.33$                 & $-0.005\pm 0.037$ \\ \hline
\end{tabular}
\end{center}
\caption{$B\to K^*\gamma$ branching fraction and direct \CP\ asymmetry measurements.}
\label{tab:kstar}
\end{table}

In addition to the already established 
$\B \to K^*\gamma$ decay, there are several known resonances that can contribute to the 
$X_s$ final state. Current measurements of higher than $K^*(892)$ mass systems are
from Belle~\cite{belle1430} and CLEO~\cite{cleo1}.
Note that the decay $\B\to \phi K \gamma$ was observed
recently by Belle~\cite{phiKgamma} for the first time. 
Theoretical predictions cover a wide range; results so far are consistent with those 
from a relativistic form factor model as in Ref.~\cite{Veseli}.
\subsection{Inclusive $b\to s \gamma$}

Two experimental approaches have been used to measure the inclusive rate 
for the $\b\to s \gamma$ process. 
The ``fully inclusive'' method measures the high energy photon 
spectrum without identifying the hadronic system $X_s$.
Continuum backgrounds are suppressed with event shape information, 
and then subtracted using off--resonance data.  
\B\ decay backgrounds are subtracted using a generic Monte Carlo 
prediction, which is cross--checked with a $b\to s\pi^0$ analysis.

\babar~\cite{BaBarbsg} has presented a preliminary result
from a fully inclusive analysis in which the ``other'' \B\ is leptonically tagged 
to almost completely suppress the continuum background.
CLEO~\cite{CLEObsg} has published a measurement combining several techniques to reduce
the background~\footnote{CLEO has recently published 
a search for baryons in $b\to s\gamma$ events~\cite{CLEObaryons}, setting upper limits on corrections
to the ${\cal B}$ and the photon energy spectrum, which are less than half of the combined quoted statistical
and systematical errors in Ref.~\cite{CLEObsg}.}.

A ``semi--inclusive'' method, which measures a sum of exclusive $B\to X_s\gamma$ 
decays, has been used by both \babar~\cite{BaBarbxs} and Belle~\cite{BELLEbxs}.
The hadronic $X_s$ system is reconstructed by \babar~(Belle) in 12 (16) final states
with a mass range up to 2.40 (2.05)~\gev. This includes about 50~\%\ of all
$b\to s\gamma$ final states. Continuum and \B\ decay backgrounds are subtracted 
by a fit to the beam--constrained \B\ mass in the same way as in an exclusive analysis.   

Figure~\ref{fig:bsgBF} summarizes the measurements of the $b\to s\gamma$ branching fraction.
The theoretical error is quoted as the extrapolation of the inclusive rate 
from the measured energy range to the full photon spectrum.
CLEO has a lower threshold (2.0~\gev) than \babar~(2.1~\gev) and Belle (2.25~\gev). 
Presently, errors are slightly larger than the theoretical uncertainty.
Computing a world average is complicated by the correlated 
systematic and theoretical errors. 
The fully inclusive method has a dominant systematic error from the \B\ decay 
background subtraction. 
The semi--inclusive method has a dominant systematic error from the 
efficiency for reconstructing the final states, including a correction for 
missing final states that are not considered. 
The average branching fraction reported, ${\cal B} = (3.40 \pm 0.39)\times 10^{-4}$, is computed assuming the
systematic errors uncorrelated, for simplicity. A first attempt to consider the correlations
among the errors can be found in~\cite{steve}.

The present ${\cal B} ( B \rightarrow X_s \gamma)$ measurements 
already provide a significant constraint on the SUSY parameter space. 
For example limits on new physics contributions to $\B \rightarrow X_s \gamma$ have been calculated using the minimal 
supergravity model (SUGRA)~\cite{hewett} and charged Higgs bosons~\cite{misiak}.

So far, only CLEO~\cite{cleo3} has measured the direct \CP\ asymmetry, using a technique which
does not suppress the background coming from $b\to d \gamma$ decays (which is expected to have a large \CP\ asymmetry). 
Thus, the measured direct \CP\ asymmetry is 
$0.965\times {\cal A}_{CP}(\B \rightarrow X_s \gamma) + 0.02\times{\cal A}_{CP}(\B \rightarrow X_d \gamma) = (-0.079 \pm 0.108 \pm 0.022) \times (1.0 \pm 0.03)$. The first error is statistical, while 
the second and third errors represent additive and multiplicative systematic 
uncertainties, respectively. The theoretical expected ${\cal B}(\B \rightarrow X_d \gamma)$ 
is used. Results are consistent with no asymmetry.
 
\babar~\cite{BaBarbxs} and CLEO~\cite{CLEObsg} have published a measurement of the photon energy spectrum
down to a threshold $E_{\gamma}^* > 2.1$ and 2.0~\gev, respectively, 
where $E_{\gamma}^*$ is measured in the \B~\footnote{Note that in case of a semi--inclusive analysis results can be 
shown in terms of the mass of the hadronic system $X_s$ or, equivalently, in terms of the photon energy, 
$E_{\gamma}$, because $E_{\gamma} = {{M^2_\B - M^2_{X_s}}\over{2M_\B}}$
in the \B\ rest frame.} and in the laboratory rest 
frame, respectively (see Figure~\ref{fig:inclusive}).

From the measured spectrum \babar\ and CLEO have extracted the first moment in
the \B\ rest frame, $\langle E_\gamma \rangle$, finding $\langle E_\gamma \rangle = 2.35 \pm 0.04 \pm 0.04$~\gev\ and 
$\langle E_\gamma \rangle = 2.346 \pm 0.032 \pm 0.011$~\gev, respectively.
Using expressions in the $\overline {MS}$ renormalization scheme, to order
$1/M_B^3$ and order $\alpha_s^2 \beta_0$~\cite{ligeti}, \babar\ and CLEO obtain 
$\bar \Lambda = 0.37 \pm 0.09 \pm 0.07 \pm 0.10$~\gev\ and $\bar \Lambda = 0.35 \pm 0.08 \pm 0.10$~\gev\ 
from the first moment.  The errors are statistical, systematic (combined in the CLEO measurement) and theoretical, respectively. 

Moreover, CLEO has used their measured $\B \to X_s \gamma$ photon energy spectrum
to determine the light--cone shape function. Using this information, CLEO extracts
$| V_{ub}| = (4.08\pm 0.34 \pm 0.44 \pm 0.16\pm 0.24)\times 10^{-3}$~\cite{CLEO-endpoint},
where the first two uncertainties are experimental and the last two from theory.

\hspace{-5cm}
\begin{figure}[hbtp]
 \begin{minipage}[c]{8cm}
\hbox to\hsize{\hss
\includegraphics[width=8cm,height=9cm]{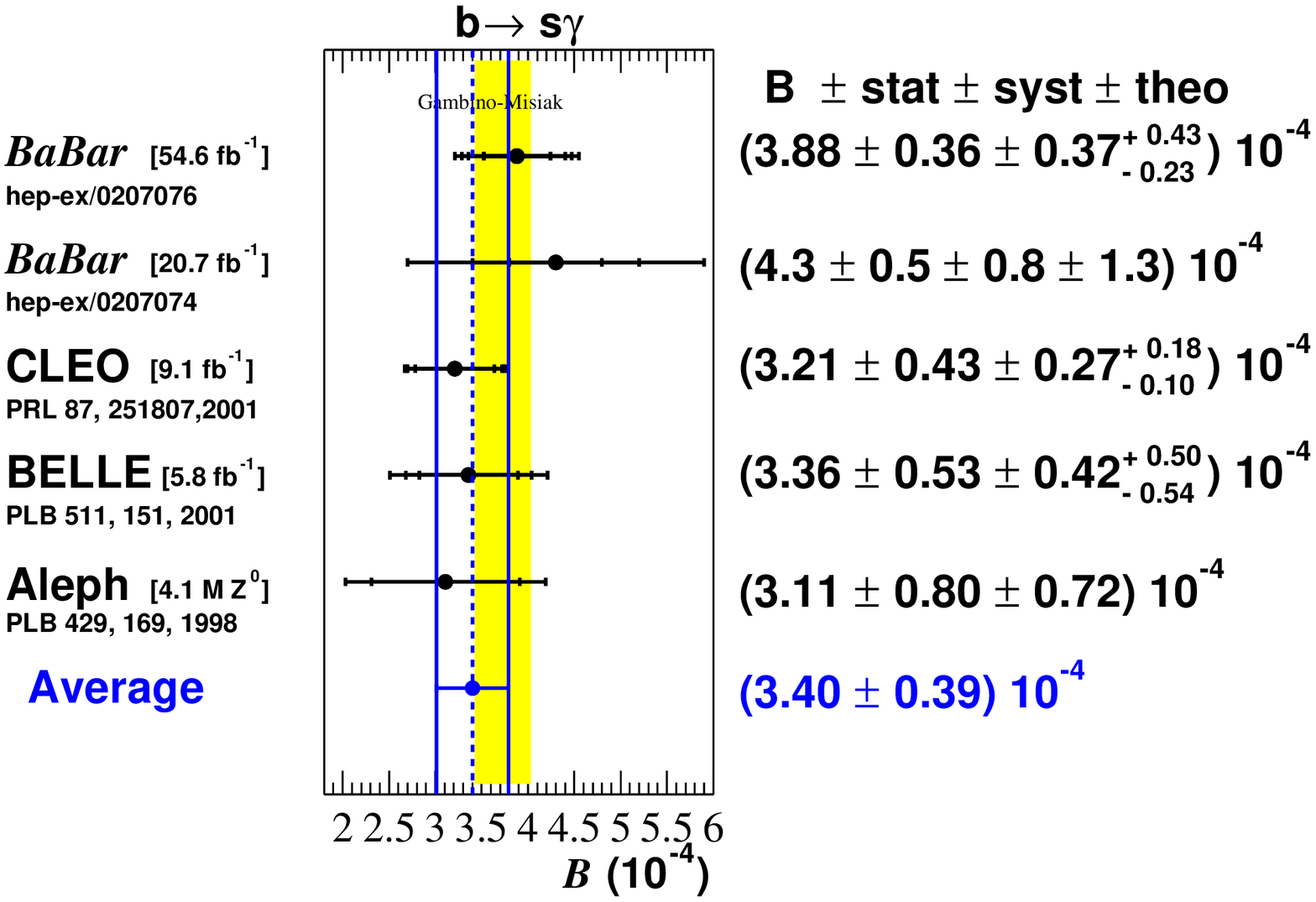}
\hss}
\caption{Summary of $b\to s\gamma$ branching fractions. The shaded band 
shows the theoretical prediction described in Ref.~\cite{misiak}.}
\label{fig:bsgBF}
\end{minipage}
\hspace{1cm}
 \begin{minipage}[c]{8cm}
\begin{tabular}{c}
\includegraphics[width=6.5cm,height=4.5cm]{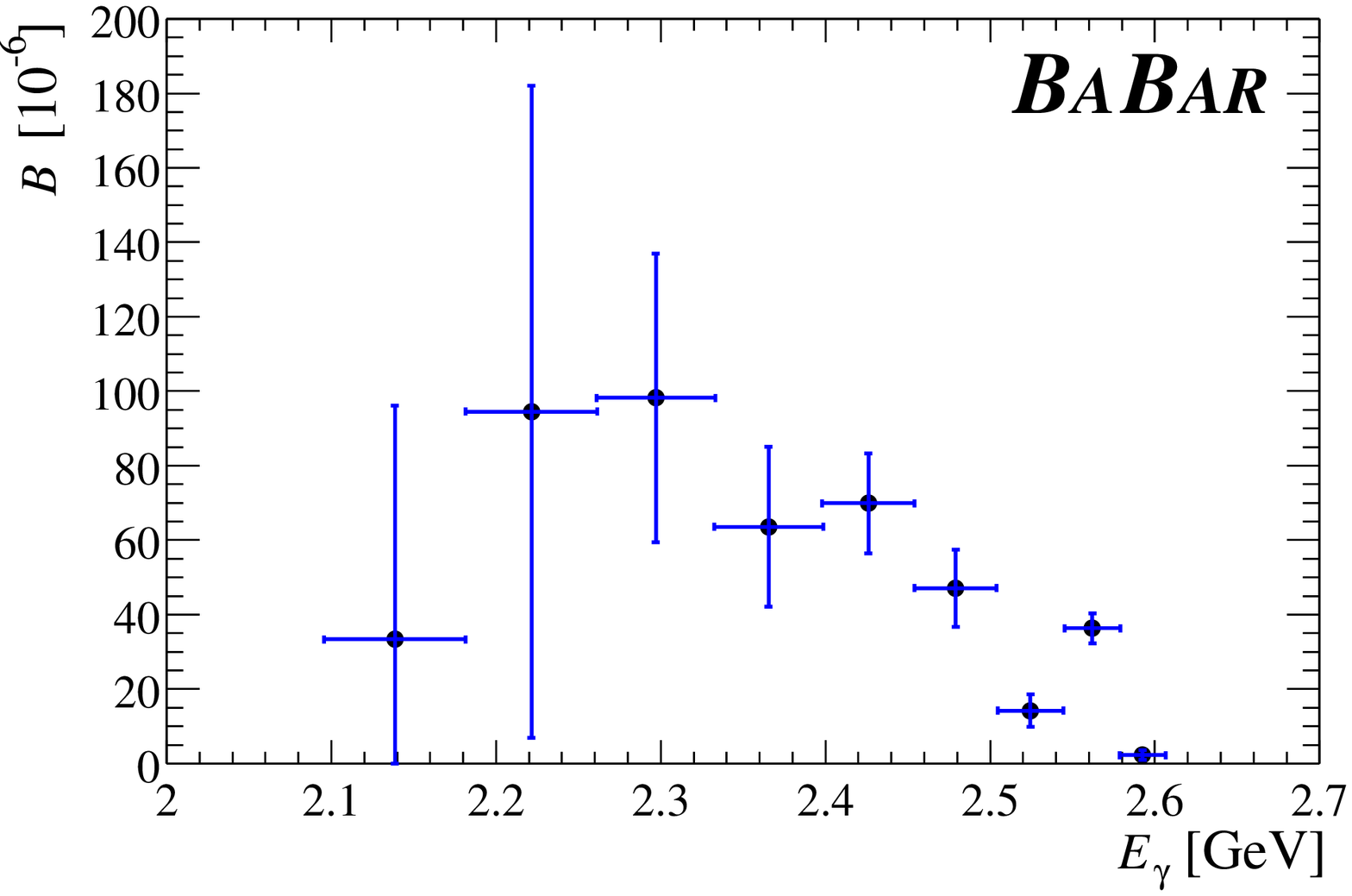}\\
\includegraphics[width=6cm,height=4.5cm]{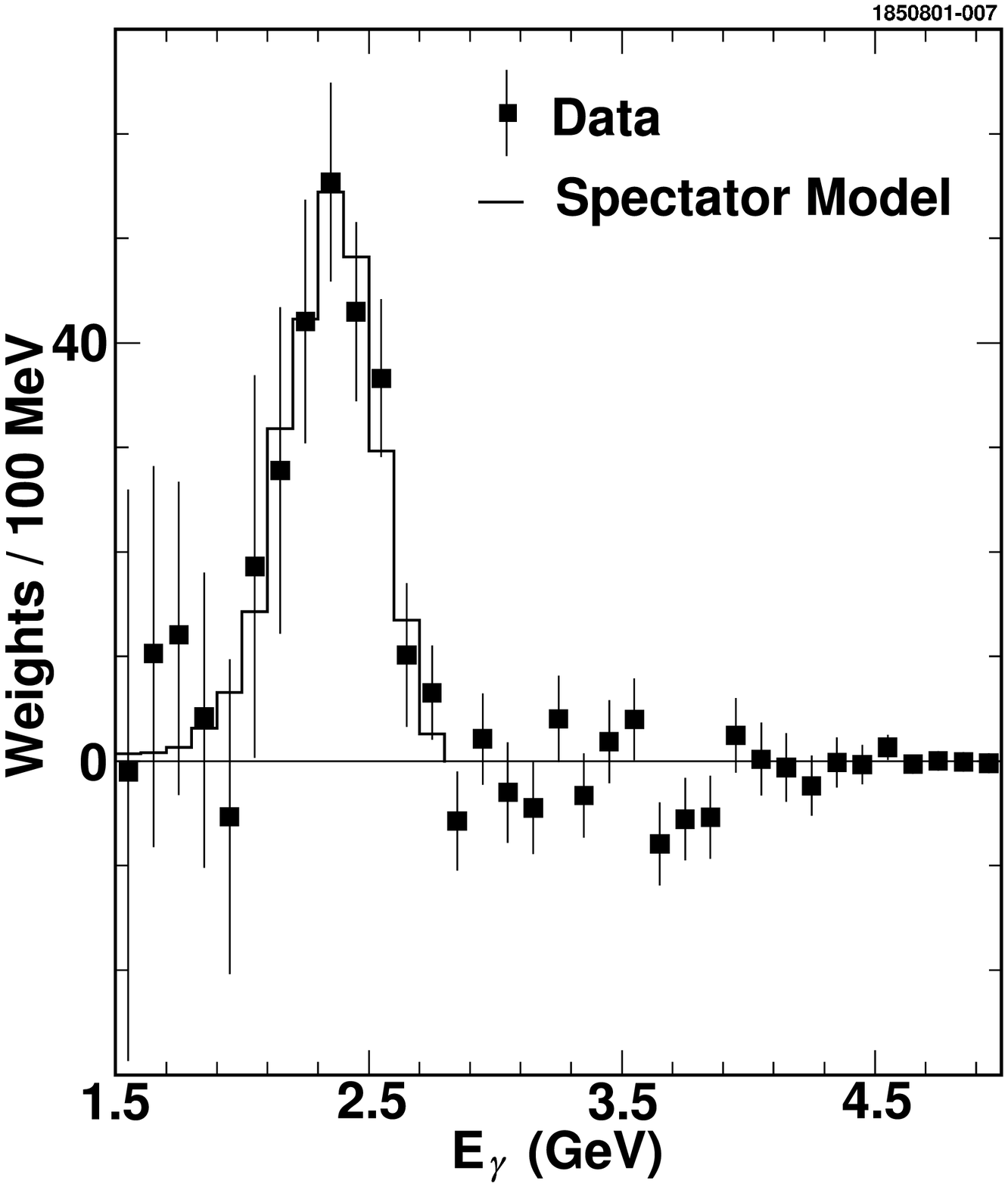}\\
\end{tabular}
\caption{\babar~\cite{BaBarbxs} (upper plot) and CLEO~\cite{CLEObsg} (lower plot) photon energy spectrum in $b\to s\gamma$
decays.}
\label{fig:inclusive}
\end{minipage}
\end{figure}
\hspace{-5cm}
\section{$b \to \d \gamma$ final states}

Both inclusive and exclusive $b \rightarrow d \gamma$ decays, which are 
suppressed by $\mid V_{td} / V_{ts} \mid^2 \sim 1/20$ with respect to corresponding
$b \rightarrow s \gamma$ modes, have not been seen yet.
An NLO calculation, which includes long--distance effects of $u$--quarks in the penguin loop,
predicts a range of
$ 6.0 \times 10^{-6} \leq {\cal B}(B \rightarrow X_d \gamma) \leq 2.6 \times
10^{-5}$~\cite{ali2} for the inclusive branching fraction. 
The uncertainty is dominated by imprecisely known CKM parameters. 

A branching fraction measurement of 
$B \rightarrow X_d \gamma$ provides a determination
of $\mid V_{td} / V_{ts} \mid$ with small theoretical uncertainties.
A determination of $\mid V_{td} / V_{ts} \mid$ in the exclusive modes
$B \rightarrow \rho(\omega) \gamma$ bears enhanced model uncertainties,
since form factors are not precisely known.

The \CP\ asymmetry predicted in the SM for the inclusive process is foreseen between $\sim 7~\%$
and $\sim 35~\%$~\cite{ali2}.

Studies of the $b\to d \gamma $ decays for now focus primarily 
on searching for the exclusive process $\B\to \rho/\omega \gamma$.
The corresponding branching fraction is predicted to be
${\cal B} (B  \rightarrow \rho \gamma) = (1.6^{+0.8}_{-0.5}) \times 10^{-5}$
\cite{bosch}, while the \CP\ asymmetry is of the order of $10\%$ \cite{bosch}.

From the experimental point of view, the $\B\to \rho(\omega) \gamma$ 
is more difficult than $\B\to K^*\gamma$ because the backgrounds are bigger since this mode is
CKM suppressed and $u \bar u, d \bar d$ continuum processes are enhanced
compared to $s \bar s$ continuum processes.

The smallest upper limits on the exclusive decays $B\to \rho(\omega)\gamma$ 
come from \babar~\cite{BABARrhogamma}, which uses a neural network to 
suppress most of the continuum background. The $B\to K^*\gamma$ events 
are removed using particle identification to veto kaons, with a $K\to\pi$ 
fake rate of $\approx$1\%. A multi--dimensional likelihood fit is made 
to the remaining events (see Figure~\ref{fig:rhogamma}) to give 90~\% C.L. upper limits 
of 1.2, 2.1 and 1.0$\times 10^{-6}$ on $\rho^0\gamma$, $\rho^+\gamma$ 
and $\omega\gamma$, respectively. Assuming isospin symmetry, this gives a 
combined limit ${\cal B}(B\to\rho\gamma)<1.9\times 10^{-6}$ (90~\% C.L.).
Limits from Belle and CLEO can be found in Refs.~\cite{BELLErhogamma} and \cite{cleo1}, respectively.

Of particular theoretical interest is the ratio ${\cal B}(\B\to\rho\gamma)$ 
to ${\cal B}(\B\to K^*\gamma)$ as most of the theoretical uncertainty
cancels and so it can be used to determine the ratio $\mid V_{td} / V_{ts} \mid$.

As described in Ref.~\cite{Alirhogamma}, the ratio can be written as:
\begin{displaymath}
{{{\cal B}(\B\to\rho\gamma)}\over{{\cal B}(\B\to K^*\gamma)}} = 
S_\rho |{{V_{td}}\over{V_{ts}}}|^2
\left({{1-m_{\rho}^2/M_B^2}\over{1-m_{K^*}^2/M_B^2}}\right)^3
\zeta^2[1+\Delta R]
\end{displaymath}
where $S_\rho$ is 1/2 (1) for $\rho^0$ ($\rho^\pm$) mesons, 
$\zeta $ is the ratio of the HQET form factors, 
$\Delta R$ accounts for possible weak annihilation 
and long distance contributions which appear mainly in $\B^+\to\rho^+\gamma$.
Eventually these can be checked by comparing $\rho^+\gamma$ and $\rho^0\gamma$.  
Using the above equation the constraint on the CKM elements $\mid V_{td} / V_{ts} \mid$ is shown in the $\rho/\eta$ plane
in Figure~\ref{fig:lunghi}. It is not as tight as the constraint from $B_s/B_d$ mixing. 
However, New Physics may appear in different ways in penguin and mixing 
diagrams, so it is important to measure it in both processes. 

\hspace{-5cm}
\begin{figure}[hbt]
 \begin{minipage}[c]{8cm}
\hbox to\hsize{\hss
\includegraphics[width=7cm,height=9cm]{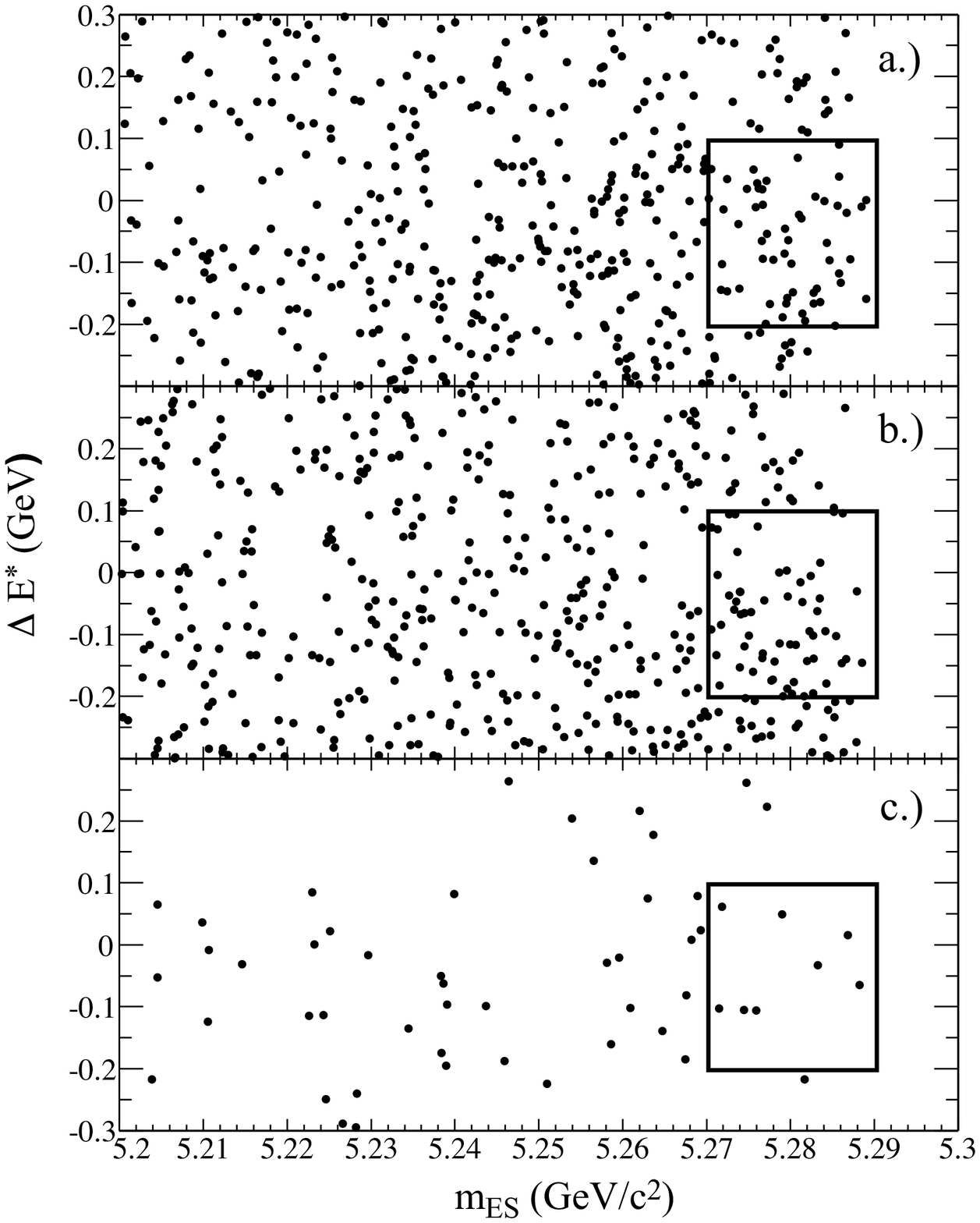}
\hss}
\caption{\babar~\cite{BABARrhogamma} $ \de$ versus the $\mes$ multi--dimensional likelihood fit regions for the  
(a) $\B \to \rho^0\gamma$ (b) $\B \to \rho^+\gamma$ (c) $\B \to \omega\gamma$ candidates.}
\label{fig:rhogamma}
\end{minipage}
\hspace{1cm}
 \begin{minipage}[c]{8cm}
\hbox to\hsize{\hss
\includegraphics[width=8cm,height=5cm]{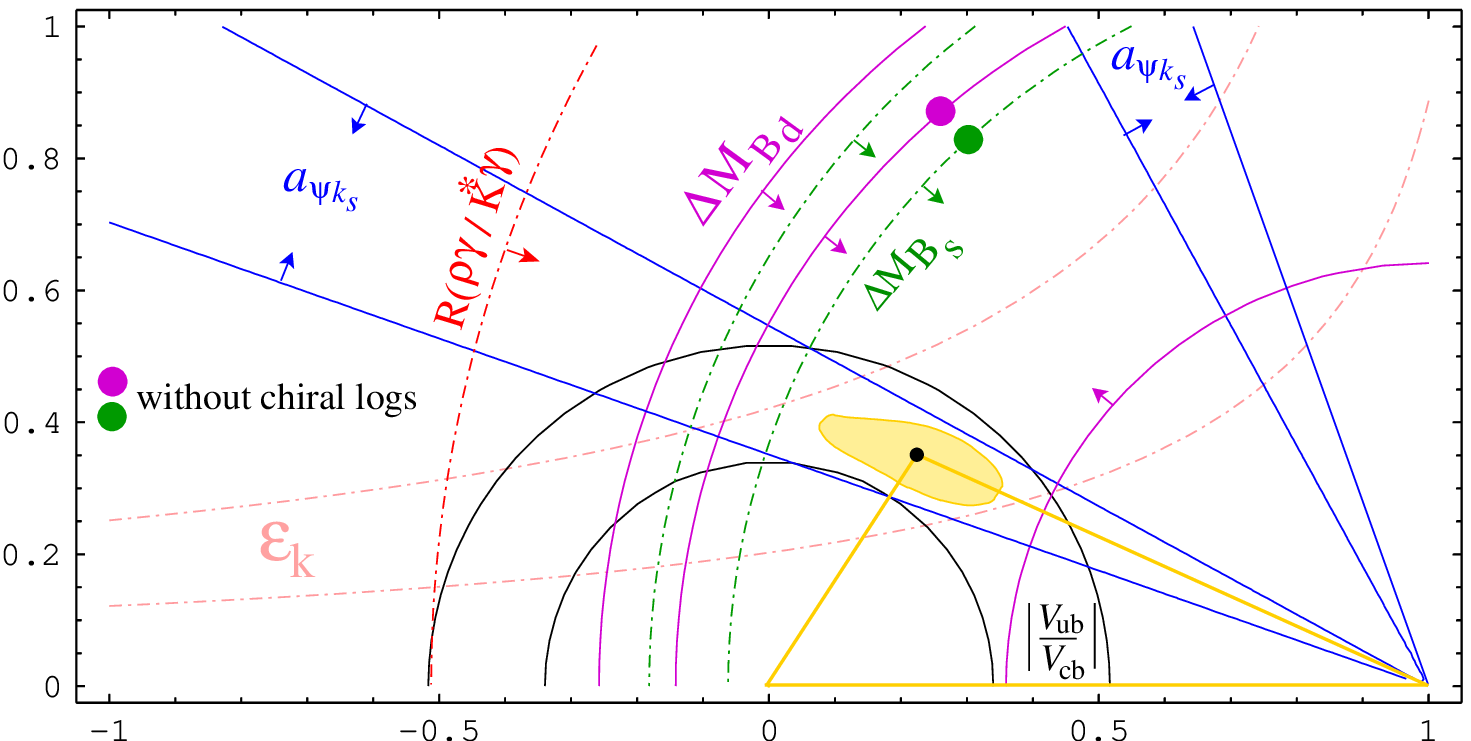}
\hss}
\caption{Constraint on CKM unitarity triangle obtained in Ref.~\cite{Alirhogamma} from the 
upper limit on $B\to\rho\gamma$~\cite{BABARrhogamma}.}
\label{fig:lunghi}
\end{minipage}
\end{figure}
\hspace{-5cm}
\section{Conclusions and Outlook}
A review of recent experimental results of radiative penguin decays $b\to s(d)\gamma $ is presented.
The $b\to s\gamma$ process in the exclusive and exclusive final states is well established
and more statistics can be used to improve the limits (or indirectly find evidence) of New Physics and
to improve the measurement of the photon energy spectrum, toward a better 
determination of the CKM matrix elements.

There is not yet evidence of $b\to d\gamma$ decays but  
\babar\ and Belle expect to collect $\approx 500~\invfb$ by 2005.
This should be sufficient to observe $B\to\rho\gamma$. 
It may also be feasible to measure the inclusive $b\to d\gamma$ rate. For the measurement of 
$|V_{td}/V_{ts}|$, the ratio of $b\to d\gamma$ 
to $b\to s\gamma$ has much smaller theoretical uncertainties 
than the ratio of the exclusive decays.

\end{document}